\documentstyle[multicol,aps,epsf]{revtex}
\def\gtwid{\mathrel{\raise.3ex\hbox{$>$\kern-.75em\lower1ex\hbox{$\sim$}}}}

\begin{document}
\title{Slowly Divergent Drift in the Field-Driven Lorentz Gas}
\author{P.~L.~Krapivsky and S.~Redner}

\address{Center for Polymer Studies and Department of Physics Boston
University, Boston, MA 02215}
\maketitle
\begin{abstract}
 
The dynamics of a point charged particle which is driven by a uniform
external electric field and moves in a medium of elastic scatterers is
investigated.  Using rudimentary approaches, we reproduce, in one
dimension, the known results that the typical speed grows with time as
$t^{1/3}$ and that the leading behavior of the velocity distribution
is $e^{-|v|^3/t}$.  In spatial dimension $d>1$, we develop an
effective medium theory which provides a simple and comprehensive
description for the motion of a test particle.  This approach predicts
that the typical speed grows as $t^{1/3}$ for all $d$, while the speed
distribution is given by the scaling form $P(u,t)=\langle
u\rangle^{-1}f(u/\langle u\rangle)$, where $u=|v|^{3/2}$, $\langle
u\rangle\sim \sqrt{t}$, and $f(z)\propto z^{(d-1)/3}e^{-z^2/2}$.  For a
periodic Lorentz gas with an infinite horizon, {\it e.\ g.}, for a
hypercubic lattice of scatters, a logarithmic correction to the
effective medium result is predicted; in particular, the typical speed
grows as $(t\,\ln t)^{1/3}$.

\end{abstract}
\begin{multicols}{2}

\section{INTRODUCTION}

At the turn of the century, Drude developed a qualitative theory for
electrical conduction in metals\cite{drude}.  To establish a more solid
basis for the Drude theory, Lorentz\cite{lorentz} suggested an idealized
model for electron transport in metals in which: (i)
electron-electron interactions are ignored, (ii) the background atoms
are considered to be immobile spherical scatters, and (iii) the
electron-atom interaction is described by elastic scattering.  This {\em
Lorentz gas}\cite{kac} has played a large role in developing our
understanding of diffusive transport in random media.

An important feature of the Lorentz gas is the independence of the
electrons.  This implies that the underlying Boltzmann equation for the
evolution of the electron velocity distribution function (VDF) is
linear.  Because of this simplification, the Boltzmann equation has
proven fruitful in understanding the properties of the Lorentz gas (see,
{\it e.\ g.},\cite{hauge} and references therein).  These investigations
have established that, under relatively general conditions, a test
particle moves diffusively and that its diffusivity can be computed in
terms of the geometric properties of the background scatterers.  The
Lorentz model is also simple enough to be amenable to rigorous
analytical studies (see, {\it e.\ g.}, \cite{bs,mz,fm,mr,bsc,bleher}).
In particular, for a periodic Lorentz gas in two dimensions with an
``infinite'' horizon ({\it i.\ e.}, there exist free trajectories of
infinite length), anomalous diffusion of the form $\langle
r^2\rangle\propto t\,\ln t$ has been proved\cite{bleher}; this anomalous
diffusion is also expected to hold in arbitrary dimension.

Paradoxically, much less is known about the problem which originally
motivated the introduction of the Lorentz gas model, {\it i.\ e.}, the
motion of a charged test particle in a scattering medium under the
influence of a spatially uniform electric field.  Lorentz himself
constructed an approximate stationary solution to the Boltzmann equation
by a perturbative expansion around the Maxwell-Boltzmann
distribution\cite{lorentz}.  From this solution, Lorentz reproduced the
basic results of the Drude theory.  Unfortunately, the starting point of
Lorentz's analysis is erroneous.  If the scattering is elastic (no
dissipation), then an electron will necessarily gain kinetic energy as
it accelerates in the field and a stationary asymptotic VDF will not
exist.  This dilemma motivated investigations of the field-driven
Lorentz gas in which some form of dissipation is explicitly
incorporated\cite{mhb,lh,ohm}, so that it is possible to obtain Ohm's
law.

In the absence of dissipation, however, Piasecki and Wajnryb\cite{pw}
were apparently the first to recognize the fundamental ramifications
that arise from the non-stationarity of the system.  From an exact
solution to the Boltzmann equation in one dimension and an asymptotic
solution for general $d$ and with the crucial assumption of collision
isotropy, they found: (i) the typical velocity, $v_{\rm rms}$, grows
with time as $t^{1/3}$, and (ii) the VDF has a non-stationary, but
symmetric asymptotic form whose controlling factor is $e^{-|v|^3/t}$.

Our goal in the paper is to develop simple and physically transparent
approaches to understand the behavior of the field-driven Lorentz gas.
We begin by considering the one-dimensional system in Sec.\ II, where we
substantially reproduce the results of Piasecki and Wajnryb\cite{pw}.
We first construct a random walk argument to explain the mechanism that
gives rise to the slow $t^{1/3}$ increase of the root-mean-square
velocity ($v_{\rm rms}$) with time.  This argument relies on the
assumption that each scattering event is spatially isotropic.  This
isotropy is a basic feature of elastic scattering from an immobile
sphere only for the physical case of three dimensions\cite{ll}.  If one
postulate isotropic scattering even in one dimension, the results
obtained can be expected to mimic three three-dimensional behavior.  The
Langevin and underlying Fokker-Planck equations for the speed
distribution function (SDF) are also investigated to obtain both the
time dependence of the typical speed and, more generally, the asymptotic
form of the SDF.  Finally, we develop a Lifshitz argument\cite{lifshitz}
to reproduce the asymptotic form of the SDF with minimal calculation.

In Sec.\ III, we study the field-driven Lorentz gas for arbitrary
spatial dimension $d$.  We argue that there are basic differences in the
field-driven Lorentz gas for $d=1$ and for $d>1$.  In greater than one
dimension, the freely-accelerated trajectory segments between scattering
events are biased by the field, leading to anisotropy in the spatial
position of the test particle at the next scattering event.  To account
for this bias in a relatively simple manner, we introduce an effective
medium theory.  In this description, a particle begins at the center of
a ``transparency'' sphere of radius equal to the mean-free path.  The
particle freely accelerates until it reaches the sphere surface.  This
defines a collision, whereupon the test particle starts at the center of
a new transparency sphere.  We will generally assume isotropic
scattering in each collision, a feature of elastic scattering from hard
spheres in three dimensions.  However, there is preferential
backscattering for For $d<3$ and preferential forward scattering for
$d>3$.  This short-range correlation apparently does not affect the
asymptotic motion of the test particle, so that we generally focus on
isotropic scattering.

For isotropic scattering, it is simple to quantify the field-induced
bias of the test particle as it moves within a transparency sphere.  A
random-walk argument of a similar spirit to that given in one dimension
indicates that the influence of the bias is of the same order as the
stochasticity caused by scattering.  This implies that the SDF will obey
one-parameter scaling.  From the solution to the underlying
Fokker-Planck equation, we find, $P(u,t)\propto
t^{-1/2}z^{(d-1)/3}\exp(-z^2/2)$, where $u=|v|^{3/2}$ is a convenient
``speed'' variable and the scaling variable $z$ is proportional to
$u/t^{1/2}$.  We also extend the effective medium theory to the case
where the mean-free path is chosen from a distribution
$\rho(\ell)\propto \ell^{-\mu}$.  This form accounts for the asymptotic
distribution of mean free paths encountered by a test particle when
(small) scattering centers are on regular lattice sites.  As might be
expected, when $\mu>3$, corresponding to a finite second moment
$\langle\ell^2\rangle$ of $\rho(\ell)$, the transport of the test
particle is nearly identical to the case where the mean-free path is
fixed.  However for $\mu< 3$, {\it i.\ e.}, for a distribution
$\rho(\ell)$ with $\langle \ell^2\rangle=\infty$, faster asymptotic
transport arises.  The borderline case of $\mu=3$ corresponds to a
lattice array of scatterers such that an infinite horizon arises, and
logarithmic corrections in the transport laws are predicted to occur.

In Sec.\ IV, we present Monte Carlo simulation results for the motion of
a test particle in the two-dimensional effective medium.  When the
radius of the transparency circle is fixed, we obtain excellent
agreement between simulation results and our theoretical predictions for
the case of isotropic scattering.  For simulations based on the correct
scattering law for hard circles in two dimensions, virtually identical
results are obtained, {\it i.\ e.}, short range antipersistence appears
to be asymptotically irrelevant.  We also consider the case of a
power-law distribution of radii for the transparency circle,
$\rho(\ell)\propto\ell^{-\mu}$.  While the physically interesting case
of $\mu=3$ corresponds to the borderline of applicability of our naive
effective medium approach, numerical results indicate transport
properties which are close to those obtained for a fixed radius
transparency circle.

In Sec.\ V, we present a brief discussion and summary.

\section{LORENTZ GAS IN ONE DIMENSION}

\subsection{Random Walk Argument for the RMS Velocity}

Consider a sizeless test particle (electron) which moves with constant
acceleration $a=eE/m$, where $e$ and $m$ are the electron charge and
mass, and $E$ is the electric field.  The electron moves in a medium of
equally spaced point scatterers with separation $\ell$.  To mimic the
behavior of a three-dimensional system with isotropic scattering, we
allow the particle to hop with equal probability to its nearest neighbor
on the left or the right after each collision.  Thus the electron
trajectory consists of freely accelerating segments which are punctuated
by isotropic scattering events.  This is simply an isotropic random
walk, but with position- and direction-dependent time increments between
successive steps.

We use this physical picture to compute the behavior of the typical
velocity as a function of time.  Energy conservation gives

\begin{equation}
{1\over 2}mv_n^2 - eEx_n = {\rm const.}
\label{econst}
\end{equation}
where $v_n$ and $x_n$ refer to the electron velocity and position
immediately after the $n^{\rm th}$ scattering event.  We rewrite this as

\begin{equation}
v_{n+1}^2-v_{n}^2 = {2eE\over m}(x_{n+1}-x_{n}) = \pm {2eE\ell\over m}.
\label{de}
\end{equation}
Because of the postulated isotropic scattering, $v_{n+1}^2-v_{n}^2$ is
equally likely to be positive or negative.  Thus we conclude that
$v_n^2$ undergoes a simple random walk as a function of $n$, with an
elementary step size given by $v_0^2\equiv 2eE\ell/m$.  As a result,

\begin{equation}
\langle v_n^2\rangle=\sqrt{n}\,v_0^2, \qquad {\rm or} 
\quad v_{\rm rms}=n^{1/4}\,v_0.
\label{vrms}
\end{equation}

To determine the dependence of $v_{\rm rms}$ on time, we write the
time increment between successive collisions as 
\begin{equation}
dt_n\equiv t_{n+1}-t_{n}\approx\ell/v_{n}.
\label{dtn}
\end{equation}
The last approximation applies when the typical speed is large so that
the acceleration between scatterings can be neglected, an assumption
which can be verified {\em a posteriori}.  The total elapsed time for
$n$ collisions is therefore
\begin{equation}
t=\sum_{k=1}^n dt_k \sim {\ell\over v_0} \int_1^n {dk\over k^{1/4}}
\sim {\ell\over v_0}n^{3/4}.
\label{time}
\end{equation}
Solving for $n$ as a function of time and substituting into
Eq.~(\ref{vrms}), gives the fundamental result
\begin{equation}
v_{\rm rms}\sim v_0\left({v_0t\over\ell}\right)^{1/3} \sim
(a^2\ell t)^{1/3}.
\label{vrmst}
\end{equation}

It is instructive to compare the time dependence of $v_{\rm rms}$ with
that of the average velocity in the field direction.  The latter can be
computed from the recursion relation

\begin{equation}
v_{n+1}\approx \pm v_{n}+adt_{n}\approx \pm v_{n}+{a\ell\over
v_{n}},
\label{vn}
\end{equation}
By isotropy, the factor of $\pm 1$ occurs equiprobably for each
scattering.  Since the typical speed grows indefinitely, we again ignore
the acceleration during the free flight between adjacent sites, so that
$v_{\rm drift} \equiv\langle v_n\rangle\approx\langle(a\ell/v_{\rm
rms})\rangle$.  As a function of time, this may be rewritten as

\begin{equation}
v_{\rm drift}(t)\sim \left({a\l^2\over t}\right)^{1/3}.
\label{vav}
\end{equation}
Thus the average drift velocity {\it decreases} with time, even though
the rms velocity grows with time.  Therefore the VDF becomes
systematically more isotropic in the long time limit\cite{pw}.

Finally, making use of Eq.~(\ref{vav}), one can estimate the average
displacement $\langle x(t)\rangle$ in the field direction to be,

\begin{equation}
\langle x(t)\rangle\sim v_{\rm drift}(t)\,t 
\sim (a\l^2 t^2)^{1/3}.
\label{xav}
\end{equation}
Alternatively, this same result follows directly from energy
conservation, Eq.~(\ref{econst}), and the time dependence of $v_{\rm
rms}(t)$ from Eq.~(\ref{vrmst}).

\subsection{The Speed Distribution}

We now derive the speed distribution function using simple approaches
which obviate the need to solve the Boltzmann equation, as given in
\cite{pw}.  First consider the Langevin equation to describe how the
typical speed depends on $n$.  Since $v_n^2$ is randomly incremented or
decremented by a fixed amount $v_0^2$ in a single collision, we may
write, in the large-$n$ limit,

\begin{equation}
{d v_n^2\over dn} = v_0^2\,\eta(n),
\label{langevin}
\end{equation}
where the noise has zero mean, $\langle\eta(n)\rangle=0$, and is
temporally uncorrelated, $\langle\eta(n)\eta(n')\rangle=\delta(n-n')$.
Since we are interested in the $n\to\infty$ limit, the continuum result
for the above correlation function is appropriate.  In this limit, the
amplitude distribution of the noise is also Gaussian.  Consequently, the
Langevin equation yields a Gaussian distribution for $v_n^2$ with a
dispersion equal to $nv_0^4$\cite{vank}.

To determine the time dependence of the speed distribution, we transform
from $n$ to $t$ by writing $dt=\ell\,dn/|v_n|$, so that
\begin{equation}
{dv_n^2\over dn}=2\ell{d|v_n|\over dt}.
\label{transform}
\end{equation}
Next, we transform the dependence of the noise correlation from $n$ to
$t$.  Writing $\delta(n-n')=\delta(t-t')\,{dt\over dn}$, gives
$\langle\eta(n)\eta(n')\rangle =\langle\eta(t)\eta(t')\rangle \ell/|v|$,
so that $\eta(n)=\eta(t)\sqrt{\ell/|v|}$.  Substituting this into the
Langevin equation, Eq.~(\ref{langevin}), gives

\begin{equation}
{d|v|\over dt}={v_0^2\over {2\ell}}\sqrt{\ell\over |v|}\eta(t),
\label{transformv}
\end{equation}
or

\begin{equation}
{d|v|^{3/2}\over dt}={3v_0^2\over 4\sqrt{\ell}}\eta(t).
\label{langevin-tform}
\end{equation}
Thus we conclude that the speed distribution function, $P(u,t)$, is
Gaussian in $u=|v|^{3/2}$, with a dispersion which is proportional to
$v_0^4/\ell$.  Then the VDF is determined from the identity
$P(v,t)dv=P(u,t)du$ to yield

\begin{equation}
P(v,t)= \sqrt{|v|\over 4\pi\ell a^2 t}\,
\exp\left[-{|v|^3\over 9\ell a^2 t}\right]
\label{1dimpvt}
\end{equation}

An independent and appealing approach to obtain the VDF is by a Lifshitz
tail argument\cite{lifshitz}.  This method is based on matching the
assumed scaling form of the VDF with the ``extreme'' contribution that
arises from a particle which is scattered in the field direction at each
collision.  This extreme tail can usually be estimated by elementary
means, and matching this to the scaling form determines the VDF.
Although this approach is heuristic, its advantages are simplicity and
wide applicability.

Our starting point is to assume that the VDF can be written in the
scaling form

\begin{equation}
P(v,t)\sim {1\over{v_{\rm rms}}} f(v/v_{\rm rms}),
\label{scaling}
\end{equation}
where the scaling function $f(z)$ is expected to approach a constant as
$z\to 0$, and vanish faster than any power law for $z\to\infty$.
Generally, this large-$z$ asymptotic behavior has the quasi-exponential
form
\begin{equation}
f(z)\sim \exp\left(-z^\delta\right),
\label{vdf}
\end{equation}
which defines the ``shape'' exponent $\delta$ of the distribution.

Let us now consider the trajectory in which the test particle is
perpetually scattered parallel to the field, so that its speed is simply
$v=at$.  Substituting into Eq.~(\ref{vdf}) and using Eq.~(\ref{vrmst})
for $v_{\rm rms}$ gives,

\begin{equation}
P(v=at,t)\sim e^{-\left(at/(a^2\ell t)^{1/3}\right)^{\delta/3}}
= e^{-(at^2/\ell)^{\delta/3}}.
\label{vdft}
\end{equation}
On the other hand, the probability $P_n$ that $n$ scattering events are
all parallel to the field equals $2^{-n}$.  For this uniformly
accelerated motion, the correspondence between $n$ and the time is
simply given by $at^2/2=n\ell$.  Thus writing $P_n$ as a function of
time and matching with the argument of the exponential in
Eq.~(\ref{vdft}), gives $\delta=3$, in agreement with the exact
solution\cite{pw}.  Parenthetically, if we define a size exponent $\nu$
through $v_{\rm rms}\sim t^\nu$, then the general scaling
relation\cite{fisher} between the size and shape exponents,
$\delta=(1-\nu)^{-1}$, fails for the field-driven Lorentz gas.

\section{LORENTZ GAS IN GREATER THAN ONE DIMENSION}

\subsection{Effective Medium Approximation}

The field-driven Lorentz gas in greater than one dimension presents a
variety of theoretical and computational challenges.  Numerical
simulations of the dissipationless system are prone to large
fluctuations and quantitative conclusions are not readily
obtained\cite{mhb}.  Because of this computational difficulty and also
because dissipation arises in any physical realization of the Lorentz
gas, simulation work has primarily focused on the field-driven system
with dissipation.  This is achieved by either allowing for
inelasticity in collision events\cite{lh}, or by introducing a
``thermostat'' which continuously extracts energy from the particle
during its free motion to maintain a constant kinetic
energy\cite{mhb,em}.  While much is known about these dissipative
systems\cite{em,cels}, our interest is in the nonstationary behavior
of dissipationless system.  In particular, we wish to understand how
the typical speed of a test particle grows with time and the form of
the resulting SDF.

\begin{figure}
\narrowtext
\epsfxsize=3in
\epsfbox{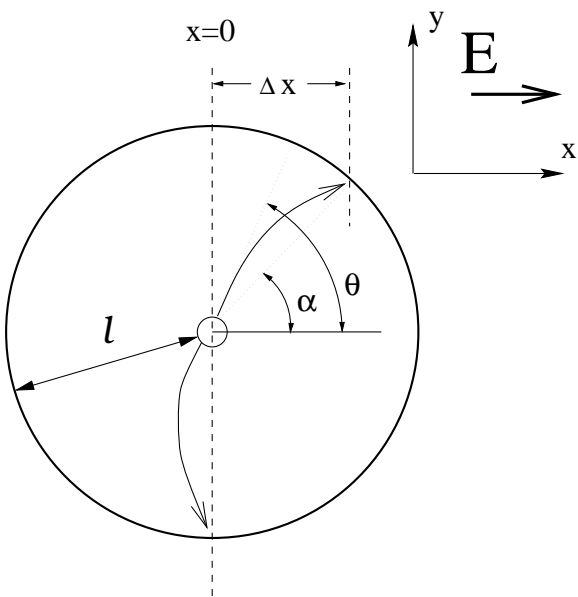}
\vskip 0.2in
\caption{``Transparency'' sphere that surrounds a scatterer.  After a 
scattering event, the test particle moves freely on a parabolic
trajectory until the next collision at the sphere boundary.  The initial
and final angular position of the test particle, $\theta$ and $\alpha$,
respectively, are indicated.  The critical trajectory is defined by the
condition that the final longitudinal position of the test particle is
at $x=0$.
\label{fig1}}
\end{figure}

Because of the inherent difficulties in describing the motion of a test
particle in a regular lattice of scatterers, we introduce an effective
medium approximation in which the true trajectory is replaced by an
effective, but physically equivalent, trajectory whose properties are
readily calculable (Fig.~1).  We assume that immediately after each
scattering event, the test particle starts at the center of a
transparency sphere of radius equal to the mean-free path $\ell$.  The
test particle then freely accelerates until the next collision when the
surface of this sphere is reached.  The collision point defines the
center of the next transparency sphere.  This construction is repeated
to generate a particle trajectory which consists of parabolic segments
(the biased free-particle motion between collisions), which are
punctuated by collision events.  We assume isotropic scattering so that
the outgoing particle direction is randomized.  This isotropy actually
occurs for hard-sphere scattering only in three dimensions.  An
elementary computation shows that there is preferential back scattering
for $d<3$ (with complete back scattering for $d=1$), and preferential
forward scattering for $d>3$.  However, this persistence for $d>3$ or
antipersistence for $d<3$ appears to be asymptotically negligible (see
below).

\subsection{The Typical Speed}

To estimate the typical speed, we need to quantify the deflection of a
trajectory during free flight.  As we shall show, this leads to an
effective bias which vanishes as the inverse square of the particle
speed.  Let us define trajectories whose collision points are in the
hemisphere $x>0$ as positively biased and {\em vice versa}.  Separating
these trajectories is a ``critical'' trajectory, in which the next
collision point is also at $x=0$ (Fig.~1).  (This critical trajectory
exists only if the initial speed satisfies $v>v_0/\sqrt{2}$; otherwise
all trajectories are deflected towards increasing $x$)~ However, since
the typical speed grows as a power law in time, the role of trajectories
in which the speed at some stage is too small to define a critical
trajectory is expected to be negligible.

By elementary mechanics, the inclination angle of this critical
trajectory is given by

\begin{equation}
\theta ={1\over 2}\sin^{-1}\left(-{a\ell\over v^2}\right)
\approx {\pi\over 2}+\left({v_0\over 2v}\right)^2, 
\quad{\rm as}\quad {v_0\over v}\to 0.
\label{bias}
\end{equation}
The longitudinal motion of the test particle can thus be viewed as a
biased one-dimensional random walk, with the bias at each step
proportional to $\epsilon\equiv (v_0/2v)^2$.  Following the same steps
as those given in Eqs.~(1) -- (6), the velocity increment between
scatterings is given by $v_{n+1}^2-v_{n}^2 = \pm v_0^2$, but with the
$\pm$ sign now occurring with respective probabilities ${1\over2}(1\pm
a\epsilon)$, where $a$ is a dimension-dependent number of order unity.
Thus, in addition to the stochastic particle motion given in
Eq.~(\ref{vrms}), a deterministic contribution also arises.  This latter
component gives, for the $n$ dependence of the speed,

\begin{equation}
\langle v_n^2\rangle\propto n\epsilon v_0^2,\qquad {\rm or}\quad 
v_{\rm rms}\propto (n\epsilon)^{1/2} v_0.
\label{biasedvn}
\end{equation}

To determine the time dependence, we relate $n$ at $t$ by using
\begin{equation}
t=\sum_{k=1}^n dt_k \propto \ell \int_1^n {dk\over (k\epsilon)^{1/2}} 
\propto {\ell\over v_0}\sqrt{n\over\epsilon}.
\label{biasedtn}
\end{equation}
Solving for $n$ as a function of $t$, substituting into
Eq.~(\ref{biasedvn}), and eliminating the factor $\epsilon$, we find
that the speed is given by
\begin{equation}
v_{\rm rms}(t)\sim v_0\left({v_0t\over\ell}\right)^{1/3}.
\label{vtav}
\end{equation}
This is the same time dependence as one dimension, where there is no
deterministic bias in the motion.  Fortuitously, the manifestations of
the isotropic scattering and the field bias are of the same order in our
effective medium theory.  This coincidence leads to a distribution of
speeds which can be described by single parameter scaling.

\subsection{The Speed Distribution}

To determine the speed distribution, we adapt our approach used in one
dimension.  First, we derive the Langevin equation for the dependence of
the typical speed on $n$, from which the underlying Fokker-Planck
equation may be written and then solved.  

In a time $\Delta\tau$ after collision $n$ (and before collision $n+1$), the
particle will be at
\begin{equation}
{\vec r} = v\,\Delta\tau\,{\hat n}+{a\,\Delta\tau^2\over 2}{\hat e}
\label{pass}
\end{equation}
with respect to the center of the transparency sphere.  Here $\hat n$
and $\hat e$ are the unit vectors in the direction of motion after the
scattering event and the electric field, respectively, {\it i. e.},
${\vec v}_n=v\hat n$ and ${\vec E}=E\hat e$.  The next collision event
takes place on the surface of the sphere ${\vec r}^{\,2} = \ell^2$.
Consequently, the time increment $\tau$ between collisions is implicitly
given by

\begin{equation}
\ell^2 = (v\tau)^2+av\tau^3(\hat n\cdot\hat e)+  
{a^2\tau^4\over 4}.
\label{t-incr}
\end{equation}

The velocity change between collisions is found from energy conservation
\begin{equation}
v_{n+1}^2-v_n^2=2a({\vec r}\cdot\hat e) 
= 2av\tau(\hat n\cdot\hat e)+(a\tau)^2.
\label{v-incr}
\end{equation}
Combining Eqs.~(\ref{t-incr}) and (\ref{v-incr}), the time increment can
be eliminated to give the analog of Eq.~(\ref{de})
\begin{equation}
v_{n+1}^2-v_n^2 
\approx 2a\ell(\hat n\cdot\hat e)+{a^2\ell^2\over v^2}
\left[1-(\hat n\cdot\hat e)^2\right].
\label{v-inc}
\end{equation}
In Eq.~(\ref{v-inc}) and below we ignore terms of order ${\cal
O}\left({v_0^6/v^4}\right)$.  The first term in Eq.~(\ref{v-inc}) is
purely stochastic, because $\langle \hat n\cdot\hat e\rangle=0$.
Since $\langle(\hat n\cdot\hat e)^2\rangle=1/d$, we may write this
stochastic term in the form $v_0^2\,\eta(n)/\sqrt{d}$.  The second term
in Eq.~(\ref{v-inc}) has both deterministic and stochastic components,
with the latter being negligible in the long time limit.  The magnitude
of the deterministic component is $(a\ell/v)^2(1-1/d)=(d-1)v_0^4/4v^2d$.
Thus we obtain the Langevin equation,

\begin{equation}
{d v_n^2\over dn} = {d-1\over 4d}{v_0^4\over v^2} + 
{v_0^2\over\sqrt{d}}\eta(n).
\label{langevin-bias}
\end{equation}
In one dimension, the deterministic term disappears
and Eq.~(\ref{langevin-bias}) coincides with Eq.~(\ref{langevin}).

Following the same steps as those given after Eq.~(\ref{langevin}), we
eliminate $n$ in favor of the time to transform the above equation to

\begin{equation}
{d |v|^{3/2}\over dt} = {3(d-1)v_0^4\over 16\ell d}{1\over |v|^{3/2}}+
{3v_0^2\over 4\sqrt{\ell d}}\,\eta(t).
\label{l-tform-bias}
\end{equation}
In this equation, the order of magnitudes of the systematic and
stochastic terms on the right-hand side are identical.  Thus
$|v|^{3/2}$ evolves by a biased random-walk process, but one in which
the bias and the dispersion are of the same scale.  This can be seen
more clearly by writing the underlying Fokker-Planck equation for
$P(u,t)$, where $u\equiv |v|^{3/2}$.  Following the standard
prescription\cite{vank}, this Fokker-Planck equation is

\begin{equation}
\label{FP}
{\partial P\over\partial t} = {9v_0^4\over 16\ell d}
\left[{\partial^2 P\over \partial u^2}
-{d-1\over 3d}\,{\partial\over \partial u}
\left({P\over u}\right)\right].
\end{equation}
We apply scaling to solve this equation.  Let us make the
scaling ansatz
\begin{equation}
\label{u-scaling}
P(u,t)={1\over \langle u\rangle}f(z)\qquad{\rm with}\quad 
z\equiv u/\langle u\rangle.
\end{equation}
Substituting Eq.~(\ref{u-scaling}) into the Fokker-Planck equation
(\ref{FP}), writing the time and velocity derivatives in terms of the
scaling variable, the partial differential equation can be separated
into two ordinary differential equations.  For the time dependence of
$\langle u\rangle$, we obtain

\begin{equation}
\label{u-eqn}
{d\over dt} \langle u\rangle^2 = {9v_0^4\over 8\ell d}.
\end{equation}
This then gives a characteristic speed which is proportional to
$(v_0^4t/\ell)^{1/3}$ or $(a^2\ell t)^{1/3}$.  For the dependence on the
scaling variable, the scaling function obeys the ordinary differential
equation
\begin{equation}
\label{f-eqn}
-f(z)-zf'(z)=f''(z)+
{d-1\over 3}\left[{f(z)\over z^2}-{f'(z)\over z}\right],
\end{equation}
where the prime denotes differentiation with respect to $z$.  One
integration gives
\begin{equation}
\label{one-int}
f'(z)=\left({d-1\over 3z}\,-z\right)f(z)+A,
\end{equation}
where $A$ is a constant.  Since $f(z)$ and its first derivative vanish
faster than any power of $z$ for $z\to\infty$, $A=0$.  The solution to
the resulting equation is 
\begin{equation}
\label{soln}
f(z)= {2^{(4-d)/6}\over{\Gamma((d+2)/6)}}\, z^{(d-1)/3}\,e^{-z^2/2}.
\end{equation}
The numerical coefficient is determined by the normalization condition
$\int_0^\infty f(z)dz=1$, $\Gamma(y)$ denotes the gamma
function\cite{AS}.  Notice that the existence of the scaling form for
the SDF hinged on the magnitude of the bias vanishing as $u^{-1}$.

\subsection{Distributed Mean-free Paths}

In both the random walk approach for $d=1$ and the effective medium
theory for $d>1$, a mean-free path which has the fixed value $\ell$ for
each scattering event was an inherent feature.  However, in the
Boltzmann equation approach of Piasecki and Wajnryb\cite{pw}, a Poisson
distribution of mean-free paths is implicitly assumed.  Moreover, there
will be a distribution of mean-free paths in any real scattering medium.
We therefore consider the physical effects that such a distribution has
on transport properties.  We consider a power law distribution of
mean-free paths,
\begin{equation}
\label{mfp}
\rho(\ell) \sim \lambda^{\mu-1}/\ell^{\mu},
\end{equation}
since this form, for $\mu=3$, corresponds\cite{bun,bd,zgnr} to the
Lorentz gas in a scattering medium with an ``infinite'' horizon ({\it
e.\ g.}, a square lattice of small scatterers).  Probability
theory\cite{gk,ba} suggests that if the distribution is relatively
sharp, the previous random walk arguments apply, while for a broad
distribution, different transport behavior arises.

Let us first consider the effect of distributed mean-free paths in one
dimension.  That is, a new mean-free path is independently chosen from
the above distribution after each scattering event.  We allow $\mu$ to
be arbitrary, since this general situation is tractable.  If the second
moment of $\rho(\ell)$ is finite, {\it i.\ e.}, $\mu>3$, then the
distribution of a sum of a large number of independent random variables,
each distributed according to $\rho(\ell)$, approaches a Gaussian and the
random walk argument of Sec.\ II applies.  In contrast, for $\mu\leq 3$,
a L\'evy distribution, whose index depends on $\mu$\cite{gk,ba},
emerges.  Making use of well-known results\cite{gk,ba} for L\'evy
distributions, we determine the $n$-dependence of $v_{\rm rms}$ to be
(the analog of Eq.~(\ref{vrms}),

\begin{equation}
v_n^2\sim v_0^2\times\cases{\sqrt{n\ln n}; &$\mu=3$,\cr
                n^{1/(\mu-1)};      &$\mu<3$,\cr}
\label{levy-vnsq}
\end{equation}
with $v_0^2=\lambda a$. 
Repeating the calculational steps in Sec.\ II, we find, for the time
dependence of $v_{\rm rms}(t)$

\begin{equation}
v_{\rm rms}(t)\sim v_0\times\cases
{\left[{v_0 t\over \lambda}\,
\ln\left({v_0t\over \lambda}\right)\right]^{1\over 3}; &$\mu=3$,\cr
\left({v_0 t\over \lambda}\right)^{1\over 2\mu-3};          
&$2<\mu<3$.\cr}
\label{levy-vrms}
\end{equation}
The average displacement in the field direction is thus given by
$\langle x(t)\rangle=v_{\rm rms}(t)^2/2a$, while the drift velocity is
$v_{\rm drift}(t)=\langle x(t)\rangle/t$.  For $\mu\leq 2$, the first
moment of $\rho(\ell)$ diverges, so that the typical mean-free path is
infinite.  Consequently, collisions become irrelevant asymptotically, so
that the typical velocity should grow linearly in time and an asymmetric
velocity distribution should arise.

\section{Numerical Simulations}

To test our theoretical predictions, we perform Monte-Carlo simulations
of particle motion in a two dimensional effective medium.  An important
element in this simulation is determining where an arbitrary parabolic
trajectory, which starts at the origin, intersects the circumference of
a concentric circle.  This involves the unwieldy solution of a quartic
equation.  However, since the typical speed grows with time, individual
trajectory segments should deviate only slightly from straight lines,
especially in the long time limit.  Thus we compute the trajectory and
the time between collisions in a perturbation series appropriate for the
large speed limit.  If the speed happens to fall below a preset
threshold such that a strongly curved trajectory segment should arise,
we impose the constraint that for this segment the particle is deflected
exactly parallel to the field.  Since the typical speed increases with
time, this ``reflecting'' boundary condition in velocity space is
anticipated to have a negligible influence on the long-time motion of
the test particle.

Our simulation algorithm therefore consists of the following basic steps
to compute the velocity and time increments between collisions.  These
steps are repeated to generate a single particle trajectory:

\smallskip

\begin{itemize}

\item If the speed is above a predetermined threshold, $v_{\rm th}$, then:

\begin{enumerate}

\item Choose a random scattering angle in the range $0\leq \theta
\leq \pi$ (see Fig.~1).

\item Determine the angular position $\alpha$ of the particle when it
hits the surface of the circle.  From elementary mechanics, the angle
$\alpha$ is perturbatively given in the large velocity limit by

\begin{equation}
\label{alpha}
\alpha=\theta - \epsilon\sin\theta +\epsilon^2\sin 2\theta +\ldots,
\end{equation}
with $\epsilon= (v_0/2v)^2$. 

\item From the angle $\alpha$, determine the change in the longitudinal
position of the particle, $\Delta x$, and thereby determine the change
in the speed of the particle by $v_f^2=v_i^2 +v_0^2\Delta x/\ell$.  Here
$v_i$ is the speed of the particle as it begins from the center of
the transparency circle, and $v_f$ is the particle speed just
before the collision at the circumference of the circle.

\item Determine the time increment, $\tau$, associated with this
trajectory.  In the large velocity limit, $\tau$ is perturbatively given
by

\begin{equation}
\label{Dt}
\tau = {1\over v}\left\{1-\epsilon\cos\theta
+{5\epsilon^2\over 2}(\cos^2\theta-1)+\ldots\right\}.
\end{equation}
Here $\tau$ and $v$ have been expressed in units of $\ell/v_0$ and
$v_0$ respectively.

\end{enumerate}

\item If the speed is less than $v_{\rm th}$, then, the scattering 
angle is taken to be $\theta=0$.  Consequently, $v_f^2=v_i^2+v_0^2$ and
$\tau=2$.  

\end{itemize}

Clearly, the sharp difference in the update of the particle motion for
initial speed smaller or larger than the threshold is a crude
approximation.  One can straightforwardly construct more accurate, but
more cumbersome rules to integrate over low-speed trajectory segments.
However, since these segments are relatively unlikely, this refinement
was not pursued.  Because of the arbitrariness in the integration of the
low-speed segments, the actual value of $v_{\rm th}$ is also somewhat
arbitrary and we chose $v_{\rm th}= v_0$.  This appears to provide a
relatively good compromise between accuracy and limiting the range over
which the arbitrary reflecting boundary condition is imposed.  To
appreciate the numerical implications, consider, for example, $v=v_{\rm
th}$.  For this case, the maximum possible deviation between the initial
and final angular position of the trajectory arises when $\theta\approx
111.5^\circ$, with $\alpha-\theta\approx -15.8^\circ$.  For $v=2v_{\rm
th}$, the maximum deviation point occurs when $\theta\approx 97^\circ$,
with $\alpha-\theta\approx -3.7^\circ$.  Thus the effect of the
curvature in the individual trajectory segments should typically be
small.

We also performed a more faithful simulation for two dimensions in which
the correct hard-circle scattering is implemented.  In place of step 1
given above, we assume that just before the $n^{\rm th}$ collision, with
incidence angle $\alpha_{n-1}$ (see Fig.~1), the test particle uniformly
illuminates the cross-section of the scatterer which is taken to be a
circle of radius $r$.  After specular reflection by the scatterer, the
difference between the incident and final angles is $d\psi
=\pi-2\sin^{-1}(b/r)$, where the impact parameter $b$ is uniformly
distributed between $\pm r$.  This angular deflection is used to compute
the outgoing angle $\theta_n=\alpha_{n-1}+d\psi$, and the corresponding
incoming angle $\alpha_n$.  Our simulation results for this more
faithful implementation of hard-circle scattering are virtually
identical to those from isotropic scattering.  Because of this
agreement, and also for simplicity, our simulations concentrated on the
case of isotropic scattering.

\begin{figure}
\narrowtext
\epsfxsize=3in
\epsfbox{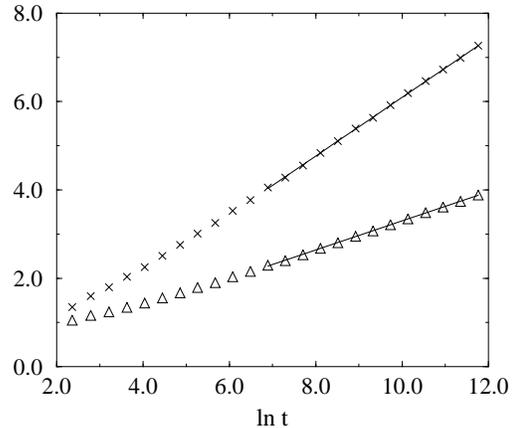}
\vskip 0.2in
\caption{Monte Carlo simulation results for 2000 walks of $1.5^{29}$
steps in a two-dimensional effective medium.  Shown are
$v_{\rm rms}(t)$ $(\Delta)$, and the mean longitudinal position $\langle
x(t)\rangle$ $(\times)$.  The straight lines represents the best fits to
the data in the range $1.5^{17}\leq t\leq 1.5^{29}$.
\label{fig2}}
\end{figure}

Typical results from this Monte Carlo simulation with isotropic
scattering are presented in Fig.~2.  Shown are $v_{\rm rms}(t)$ and
$\langle x(t)\rangle $ on a double logarithmic scale based on 2000
trajectories of $1.5^{29}\approx 127,834$ steps for the case where the
transparency circle has a fixed radius.  After some transient behavior,
the data for $t\gtwid 500$ appear to be linear and a linear
least-squares fits yields the respective slopes of 0.329 and 0.665, in
excellent agreement with the respective theoretical predictions of 1/3
and 2/3.  For true hard-circle scattering in two dimensions, our
simulations gave the corresponding exponent estimates of 0.330 and
0.662.  Thus the effect of the antipersistence in the particle motion
truly appears to be irrelevant.  In Fig.~3, we present corresponding
results for the distribution of $u=|v|^{3/2}$ at $t=1.5^{20}$ ($\circ$),
$t=1.5^{23}$ ($\diamond$), $t=1.5^{26}$ ($\nabla$), and $t=1.5^{29}$
($+$).  The raw data has been scaled so that the abscissa is
$z=u/\langle u\rangle$, while the ordinate is $f(z)=\langle u\rangle
P(u,t)$.  This scaled data at later times has then been smoothed by
averaging over a small neighborhood to reduce fluctuations.  These data
compare well with the theoretical prediction
$f(z)=(2^{1/3}/\Gamma(2/3))\times z^{1/3}\,e^{-z^2/2}$ (Fig.~3).

\begin{figure}
\narrowtext
\epsfxsize=3in
\epsfbox{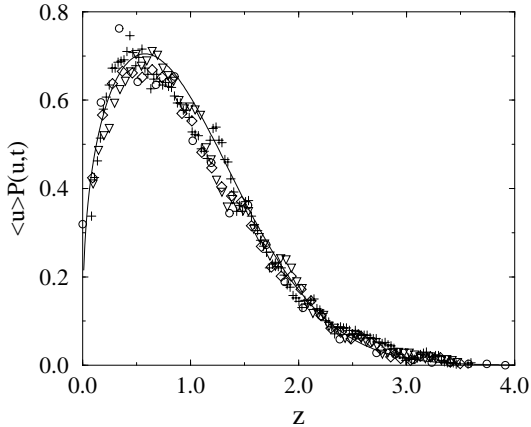}
\vskip 0.2in
\caption{Scaled distribution $\langle u\rangle P(u,t)$ versus the
variable $z\equiv u/\langle u\rangle$, where $u=|v|^{3/2}$.
Representative data shown include $t=1.5^{20}$ $(\circ)$, $t=1.5^{23}$
($\diamond$), smoothed over a 3-site neighborhood, $t=1.5^{26}$
($\nabla$), smoothed over a 5-site neighborhood, and $t=1.5^{29}$ ($+$),
smoothed over a 7-site neighborhood.  The curve is the theoretical
prediction $0.930\ldots\times z^{1/3}e^{-z^2/2}$.
\label{fig3}}
\end{figure}

As discussed previously, a lattice array of scatterers leads to a
power-law distribution of mean-free paths.  We therefore also performed
simulations of the effective medium where the radius of the next
transparency circle is chosen from the distribution $\rho(\ell)\sim
\lambda^{\mu-1}/\ell^\mu$, with $\mu=3$.  In this case, we found that
the time dependence of $v_{\rm rms}(t)$ and $\langle x(t)\rangle$ is
quite close to that obtained for the case of a fixed-radius transparency
circle.  The data for $t\gtwid 1000$ appear to be linear on a double
logarithmic scale, and a linear least-squares fits in this range yields
the respective slopes of 0.340 and 0.671 (Fig.~4).  For this case, however, the
data for $v_{\rm rms}(t)$ and $\langle x(t)\rangle$ exhibit a slight downward
trend, a feature which could be attributed to a logarithmic correction.
However, our data are insufficient to  test for such a
correction quantitatively.

\begin{figure}
\narrowtext
\epsfxsize=3in
\epsfbox{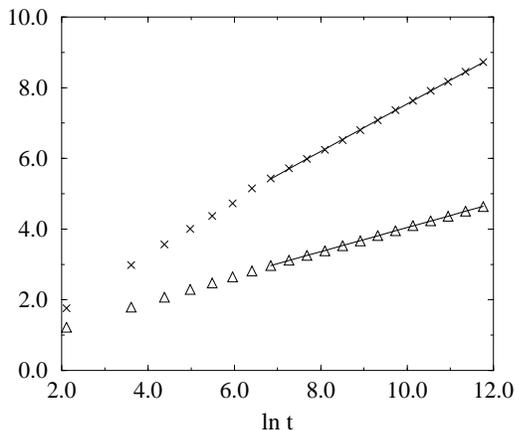}
\vskip 0.2in
\caption{Monte Carlo simulation results for 2000 walks of $1.5^{29}$
steps in a two-dimensional effective medium in which the radius $\ell$
of the transparency sphere is drawn from the distribution
$\rho(\ell)\propto \ell^{-3}$.  Shown are $v_{\rm rms}(t)$ $(\Delta)$,
and the mean longitudinal position $\langle x(t)\rangle$ $(\times)$.
The straight lines represents the best fits to the data in the range
$1.5^{17}\leq t\leq 1.5^{29}$.
\label{fig4}}
\end{figure}
The distribution of speeds also exhibits relatively good data collapse,
but there are quantitative discrepancies between the shape of the
scaling function and the prediction $f(z)\approx 0.930\ldots\times
z^{1/3}\,e^{-z^2/2}$ that fit the data for the case of a fixed-radius
transparency sphere (Fig.~5)

\begin{figure}
\narrowtext
\epsfxsize=3in
\epsfbox{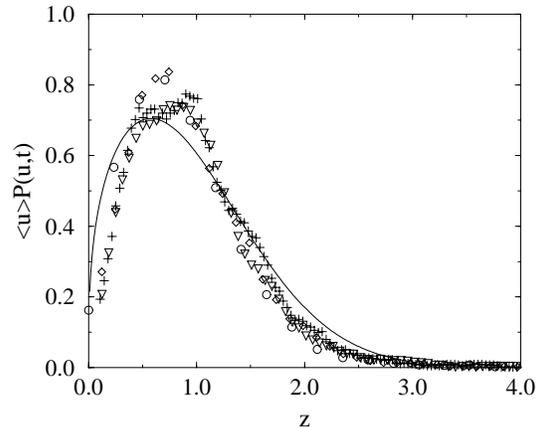}
\vskip 0.2in
\caption{Scaled distribution $\langle u\rangle P(u,t)$ versus the
variable $z\equiv u/\langle u\rangle$, where $u=|v|^{3/2}$.  The radius
$\ell$ of the transparency sphere is drawn from the distribution
$\rho(\ell)\propto \ell^{-3}$.  Representative data shown include
$t=1.5^{20}$ $(\circ)$, $t=1.5^{23}$ ($\diamond$), smoothed over a
3-site neighborhood, $t=1.5^{26}$ ($\nabla$), smoothed over a 5-site
neighborhood, and $t=1.5^{29}$ ($+$), smoothed over a 7-site
neighborhood.  The curve is $0.930\ldots\times z^{1/3}e^{-z^2/2}$.
\label{fig5}}
\end{figure}

\section{DISCUSSION AND SUMMARY}

We have investigated the transport of a charged particle which is driven
by a constant field in a dissipationless elastic and isotropic
scattering medium -- the field-driven Lorentz gas.  A fundamental and
intriguing feature of this system is that the transport is
non-stationary.  Namely, as a function of time, the typical velocity
grows as $t^{1/3}$.  Although this growth is unbounded, it is
significantly slower than a linear time dependence that would occur in
the absence of scattering.  In one dimension, we have developed a random
walk approach in which the particle hops to the right or left with equal
probability at each scattering event.  The time increment associated
with each hop between neighboring scatterers is position and direction
dependent, a feature which underlies the anomalous time dependence of
the typical velocity and mean displacement.

Based on this random walk picture, we also obtained the velocity
distribution by first writing the Langevin equation for the evolution of
the typical velocity and the underlying Fokker-Planck for the velocity
distribution. We also constructed a Lifshitz tail argument which
reproduced the correct behavior for the velocity distribution.  The
solution to the Fokker-Planck equation yields the Gaussian distribution
in the variable $u=|v|^{3/2}$,

\begin{equation}
\label{sum-udf}
P(u,t)\propto {1\over\sqrt{t}}e^{-u^2/t},
\end{equation}
which, when written in terms of $|v|$, becomes

\begin{equation}
\label{sum-vdf}
P(v,t)\propto \sqrt{|v|\over t}e^{-|v|^3/t}.
\end{equation}
Interestingly, this is similar, but not coincident with the asymptotic
velocity distribution function

\begin{equation}
\label{sum-vdf-pw}
P(v,t)\propto {1\over t^{1/3}}e^{-|v|^3/t}.
\end{equation}
obtained by Piasecki and Wajnryb\cite{pw} from the Boltzmann equation
approach.  However, this approach implicitly assumes an ``annealed''
medium in which there is a Poisson distribution of distances between
collision events.  Thus, while our random walk and the Boltzmann
approach are expected to give the same scaling of the typical speed with
time, the form of the velocity distribution from the two approach should
not be expected to coincide.  

Our random walk argument can also be applied to the interesting case of
an alternating electric field $E(t)=E_0\sin(\omega t)$ to give the
counterintuitive result that the combination of an AC field and
isotropic scattering leads to unbounded growth in the speed.  This
growth arises precisely because of the isotropy in the scattering
events.  When the time between collisions becomes less than the time for
the field to reverse, then the direction of the field becomes
irrelevant.  Consequently, our random walk argument for a DC field
directly applies and $v_{\rm rms}$ grows without bound in time.  The
validity of this statement depends only on the existence of a
well-defined typical magnitude for the field.  Thus for an AC external
electric field, the scatterers assist in the absorption of field energy
by the test particle.  In contrast, in an AC field with no scattering, a
test particle merely follows the field and the typical speed is bounded.

In higher dimensions, we introduced an effective medium approximation
which provides a simple and physically appealing description for the
motion of a charged test particle.  This approximation posits that the
test particle moves a fixed radial distance along a parabolic
field-biased trajectory within a ``transparency'' sphere and that an
isotropic collision event occurs when the particle reaches the surface
of this sphere.  The assumption of scattering when a particle moves a
fixed radial distance implies an annealed medium.  Thus one might expect
a relatively close connection between the effective medium and the
Boltzmann equation approaches.  However, because of the bias in the
free-particle trajectory segments, an initially isotropic distribution
of outgoing particle directions immediately after one scattering event
becomes anisotropic at the next scattering.  Within an equivalent
one-dimensional random walk description of the test particle motion,
this anisotropy can be described in terms of an effective bias which is
proportional to $1/v^2$.  The logical consequences of this feature again
leads to a typical speed which again grows as $t^{1/3}$, just as in one
dimension.

The effect of the field-induced bias is more pronounced in the behavior
of the speed distribution, however.  Following a similar approach as
that given for one dimension, the solution to the Fokker-Planck equation
for $u=|v|^{3/2}$ is

\begin{equation}
\label{sum-udf-gend}
P(u,t)\propto 
{1\over\sqrt{t}}\left({u\over\sqrt{t}}\right)^{(d-1)/3}e^{-u^2/t},
\end{equation}
which, when written in terms of $|v|$ gives

\begin{equation}
\label{sum-vdf-gend}
P(v,t)\propto {|v|^{d/2}\over t^{(d+2)/6}}\,e^{-|v|^3/t},
\end{equation}
while the corresponding result of Piasecki and Wajnryb is
\begin{equation}
\label{sum-vdf-pw-gend}
P(v,t)\propto {|v|^{d-1}\over t^{d/3}}e^{-|v|^3/t}.
\end{equation}
While these two forms agree for $d=2$, the coincidence is unexpected.
The Boltzmann approach explicitly builds in isotropy in the collision
events and in the intervening particle motion, while the effective
medium explicitly accounts for the field-induced bias between scattering
events.

An attractive aspect of the effective medium approach is that it can be
easily generalized to a distribution of mean-free paths, a feature which
arises in a lattice realization of the Lorentz gas.  Such a distribution
may be accounted for by a power-law distribution of sphere radii
$\rho(\ell)\propto\ell^{-\mu}$, with $\mu=3$.  This represents a
marginal case between the regime where distributed radii appear to have
no effect, for $\mu>3$, to the case where the scaling of the mean speed
with time is affected, for $2<\mu<3$.  Our numerical simulations
indicate that the case of $\mu=3$ leads to behavior similar to that of
no dispersion in the sphere radii.  However, the applicability of either
the Boltzmann equation approach or our effective medium description to a
lattice realization of the Lorentz gas has yet to be tested.

\bigskip

We thank R. S. Chivukula for a helpful discussion and J. Machta for
particularly useful advice about hard-sphere scattering and related
suggestions.  This research was supported in part by the NSF (grants
DMR-9219845 \& DMR-9632059), and by the ARO (grant DAAH04-93-G-0021).

\end{multicols} 

\end{document}